\def\1{\'{\i}}
\def\beq{\begin{equation}}
\def\eeq{\end{equation}}
\def\bea{\begin{eqnarray}}
\def\eea{\end{eqnarray}}
\def\bed{\begin{displaymath}}
\def\eed{\end{displaymath}}

\def\gae{{> \atop \sim}}

\documentclass[twocolumn,pre,aps]{revtex4}
\usepackage{graphics,graphicx}

\begin{document}

\title{Validity of numerical trajectories in the synchronization transition of complex systems}

\author{R. L. Viana $^1$, C. Grebogi $^2$, S. E. de S. Pinto $^1$, S. R. Lopes $^1$, A. M. Batista $^3$, and J. Kurths $^4$}
\affiliation{1. Departamento de F\'{\i}sica, Universidade Federal do Paran\'a, 81531-990, Curitiba, Paran\'a, Brazil. \\
        2.  Instituto de F\'{\i}sica, Universidade de S\~ao Paulo, 05315-970, S\~ao Paulo, S\~ao Paulo, Brazil. \\
        3. Departamento de Matem\'atica e Estat\'{\i}stica, Universidade Estadual de Ponta Grossa, Ponta Grossa, Paran\'a, Brazil \\
        4. Institut f\"ur Physik, Universit\"at Potsdam, PF 601553, D-14415, Potsdam, Germany
.
}

\begin{abstract}
We investigate the relationship between the loss of synchronization and the onset of shadowing breakdown {\it via} unstable dimension variability in complex systems. In the neighborhood of the critical transition to strongly non-hyperbolic behavior, the system undergoes on-off intermittency with respect to the synchronization state. There are potentially severe consequences of these facts on the validity of the computer-generated trajectories obtained from dynamical systems whose synchronization manifolds share the same non-hyperbolic properties.  
\end{abstract}

\maketitle
There is a steadfast interest in spatially extended dynamical systems, as coupled map lattices, in which both space and time are discrete variables, due to their various physical and biological applications \cite{kaneko}. Complete or identical synchronization (CS) occurs when all the coupled maps have the same value for their dynamical variables \cite{kurthsbook}. Weakly coupled maps are not usually identically synchronized, but as the coupling strength increases they can reach CS. The approach to this state occurs in an intermittent fashion, alternating quiescent states of synchronized behavior with irregular bursts \cite{pecora}. As the average duration of these quiescent states tends to infinity, we attain a CS state. The onset of the transition from CS to intermittent synchronization indicates that the dynamics of the synchronized state ceases to be hyperbolic through a mechanism called unstable dimension variability (UDV) \cite{udv}. 

A dynamical system whose invariant set (like a chaotic attractor) exhibits UDV fails to have the splitting between stable and unstable manifolds which is consistent with the dynamics. The consequences of UDV are potentially dangerous from the point of view of the reliability of the numerical trajectories one gets from the dynamical system \cite{sauer}. A computer-generated chaotic trajectory is unavoidably subjected to one-step errors caused by using finite-precision arithmetics, which makes the noisy trajectory to exponentially diverge from the true chaotic orbit it is intended to represent . However, when the dynamical system presents the shadowing property, there exists a true chaotic trajectory which stays uniformly close to the numerical trajectory for a certain time interval \cite{shadows}. This shadowing trajectory may not be that particular orbit we have been looking for, but for many purposes - as when computing statistical quantities - they can be equally useful \cite{kurths}. 

In this communication, we aim to harness the connection between the intermittent transition from CS and the shadowing breakdown induced by UDV, so as to explain the intermittency features from the statistical properties of the appropriate finite-time Lyapunov exponents. Moreover, if the shadowing breakdown is so severe that we could not find a true chaotic trajectory that stays close to the numerical trajectories we generate, the latter have their validity compromised. As we will see, this will be typically the case in the transitional regime from CS. Previous papers \cite{lai99} have dealt with shadowing breakdown {\it via} UDV in lattices of maps and oscillators with local coupling. On the other hand, the present work deals with {\it non-locally coupled} maps, for which a transition to CS is observed as the effective coupling range is varied, from a global to a local interaction form \cite{betax}.

We examine a one-dimensional chain of $N$ coupled chaotic logistic maps at the outer crisis, $x \mapsto f(x)= 4 x (1-x)$, where $x_n^{(i)} \in [0,1]$ represents the state variable for the site $i$ $(i = 1, 2, \ldots N)$ at time $n$, with a coupling prescription in which the interaction strength between sites decays in a power-law fashion with the lattice distance \cite{sandro}
\beq
\label{cml}
x_{n+1}^{(i)} = (1 - \epsilon) f(x_n^{(i)}) + \frac{\epsilon}{\eta(\alpha)} \sum_{j=1}^{N'} \frac{f(x_n^{(i+j)}) + f(x_n^{(i-j)})}{j^\alpha}
\eeq
\noindent where $0 \leq \epsilon \leq 1$ and $\alpha \geq 0$ are the coupling strength and effective range, respectively, and $\eta(\alpha) = 2 \sum_{j = 1}^{N'} j^{-\alpha}$, with $N' = (N - 1)/2$, for $N$ odd. This coupling prescription allows one to pass continuously from a local Laplacian-type coupling (when $\alpha \rightarrow \infty$), to a global ``mean-field'' coupling ($\alpha = 0$). Periodic boundary conditions and random initial conditions are assumed. 

A numerical diagnostic of CS is provided by the order parameter 
\beq
\label{order}
z_n = R_n \exp\left(2\pi i \varphi_n\right) \equiv \frac{1}{N} \sum_{j=1}^{N} \exp(2\pi i x^{(j)}_n),
\eeq
\noindent where $R_n$ and $\varphi_n$ are the amplitude and angle, respectively, of a centroid phase vector, for a one-dimensional chain with periodic boundary conditions \cite{sandro}. A time-averaged amplitude ${\bar R}$ is computed over an interval large enough to warrant that the asymptotic state has been achieved by the lattice. A CS state implies ${\bar R} = 1$, for all the phase vectors gyrate together with the same phase. When the maps are completely non-synchronized, on the other hand, we get a pattern in which the site amplitudes $x^{(j)}_n$ are so spatially uncorrelated that they may be considered essentially as random variables, such that ${\bar R}$ vanishes.
\begin{figure}
\includegraphics[width=0.8\columnwidth,clip]{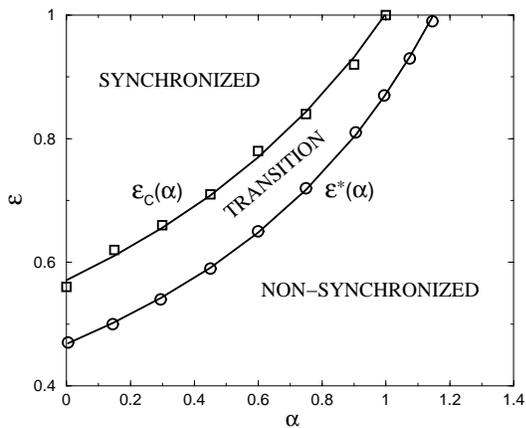}
\caption{\label{synchro} Synchronized regimes in the parameter plane of coupling strength $\epsilon$ {\it versus} effective range $\alpha$, for a lattice with $N = 21$ coupled logistic maps.}
\end{figure}

Figure \ref{synchro} represents a portion of the coupling parameter plane (strength {\it versus} effective range) exhibiting regions for which the coupled maps have (i) synchronized chaotic orbits (CS); (ii) a transitional regime where CS is eventually attained after a transient characterized by intermittent bursts of non-synchronized behavior; and (iii) completely non-synchronized orbits. These regions are bounded by the curves $\epsilon^*(\alpha)$ and $\epsilon_c(\alpha)$. The former correspond to the values of $\epsilon^*$ and $\alpha$ for which the average order parameter magnitude ${\bar R}$ ceases to be equal to the unity. The latter curve was computed by means of the finite-time Lyapunov exponents, as will be explained later on in this communication. 

Let us fix the effective range parameter at an intermediate value, say $\alpha = 0.4$, and decrease the coupling strength $\epsilon$ from its maximum value, $1.0$, to zero. For $\epsilon$ large enough we have ${\bar R}=1$, or a chaotic CS state. When $\epsilon = \epsilon_c(0.4) \approx 0.69$ this CS state starts to be interrupted by intermittent bursts of non-synchronized behavior, but eventually the stationary CS state is achieved again. The bursting becomes more frequent as the coupling strength is further decreased and, at $\epsilon = \epsilon^*(0.4) \approx 0.57$, the order parameter vanishes and the lattice becomes completely non-synchronized, never to return to a CS state again. Hence, the interval $\epsilon_c < \epsilon < \epsilon^*$ characterizes a transition region for which the intermittent bursting is a transient phenomenon.

This scenario is robust and present for a wide portion of the coupling parameter plane. As we increase the value of the effective range parameter $\alpha$ from zero, it turns out that the interval characterizing a transition region is pushed towards higher values of the coupling strength $\epsilon$. We can understand this fact since the higher the effective range $\alpha$ is, the closer we are to a locally coupled lattice, in which only the nearest neighbors contribute in a significant way, such that it becomes increasingly more difficult to have a stationary CS. We call $\alpha_c \approx 0.940$ the range parameter value for which the first critical curve reaches the upper limit given by $\epsilon_c(\alpha_c) = 1$. Accordingly, $\alpha^* \approx 1.146$ is the value where the second critical curve is such that $\epsilon^*(\alpha^*) = 1$. Hence, for $\alpha > \alpha^*$, we will not observe a stationary CS state, irrespective of how strong the coupling may be, i.e. the intermittent bursting continues for an arbitrarily long time.

Near the second critical curve in Fig. \ref{synchro} the average duration of the laminar regions obeys a power-law scaling $<\tau> \sim {(\alpha - \alpha^*)}^{-\gamma}$, when $\alpha \rightarrow {\alpha^*}^-$, and with $\gamma = 1/2$ within the numerical accuracy. This suggests that the transition to synchronization occurs through a crisis \cite{pikowsky}. The rationale for this analogy identifies a burst between laminar regions as a chaotic transient resulting from the collision of a chaotic attractor with an unstable periodic orbit. The average chaotic transient length in one-dimensional maps, like the logistic map $f(x) = ax(1-x)$ at $a=4$, obeys an identical scaling and with the same critical exponent \cite{crise}. This crisis is mediated through an unstable-unstable pair bifurcation at those points at which there is such a collision \cite{viana}. 

\begin{figure}
\includegraphics[width=0.8\columnwidth,clip]{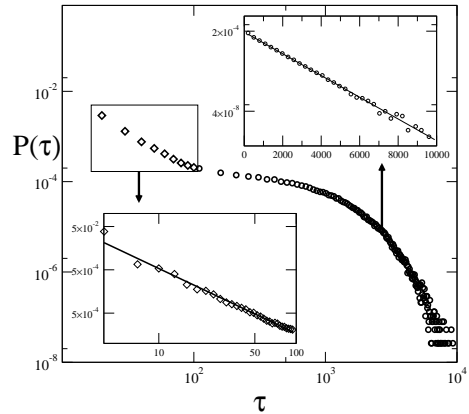}
\caption{\label{histo}Normalized histogram for the relative number of synchronization plateaus with a given length, for $\epsilon = 1.0$ and $\alpha = 1.147$.}
\end{figure}

Another distinctive feature of the intermittent transition to synchronization is the universal character of the statistical distribution of the laminar regions. In Figure \ref{histo} we present a histogram for the number of the laminar regions with respect to their lengths. Two different regimes are highlighted in the insets: for small laminar regions the histogram is well fitted by a power-law $\tau^{-\varpi}$, with $\varpi \approx 1.5$; whereas the scaling is exponential $e^{- \kappa \tau}$ for large  intervals, with $\kappa \approx 10^{-3}$. The presence of the $3/2$-scaling plus the exponential decay shows that the intermittent transition to CS is of the on-off type \cite{heagy}. We note that, in the neighborhood of the unstable-unstable pair bifurcation which mediates the crisis, a given map of the lattice chain receives random kicks through the coupling with its neighbors. These kicks, in the vicinity of the bifurcation point, are responsible for the observed on-off intermittency. In fact, the existence of two distinct scalings in Fig. \ref{histo} with a shoulder indicates the presence of noise in the on-off intermittent scenario, with a crossover time related to the noise level \cite{onoff2}. 

In the transition region, trajectories off but very near the synchronization manifold experience intermittent bursting, such that numerical diagnostics using an insufficiently large time interval could erroneously point out CS \cite{chansong}. On the other hand, this intermittent bursting is directly related to the shadowing breakdown of chaotic trajectories in the synchronization manifold of the coupled map lattice $x_n^{1} = \cdots = x_n^{N}$ \cite{pinto}. The intermittent transition to CS initiates through an unstable-unstable pair bifurcation occurring on the synchronization manifold. The unstable periodic orbit undergoing this bifurcation, and all its infinite pre-images, increase their unstable dimension by one unit. Since there remains an infinitely large number of periodic orbits that did not undergo this increase, the synchronization manifold fails to be hyperbolic through unstable dimension variability \cite{viana}. 

\begin{figure}
\includegraphics[width=0.7\columnwidth,clip]{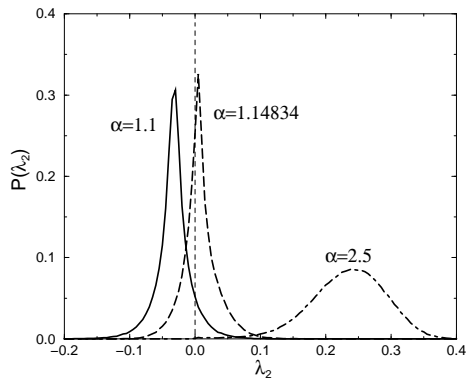}
\caption{\label{ftle}Normalized histogram of the largest transversal finite-time Lyapunov exponents for different values of the effective range $\alpha$ and $\epsilon = 1.0$.}
\end{figure}

This violates one of the mathematical requirements for hyperbolicity and shadowing, which is the continuous splitting of stable and unstable invariant subspaces along the attractor \cite{udv}. If we start off but very near the synchronization manifold, after it has lost hyperbolicity through UDV, the corresponding trajectory approaches unstable orbits with different unstable dimensions. As a result, there are both time intervals for which the trajectory is, on average, approaching the manifold, and other intervals for which it is repelled, also on average. This behavior is quantified by the largest finite-time transversal Lyapunov exponent, $\lambda_2(n)$ \cite{betax}. If it is negative (positive) a trajectory close to the CS state will be attracted (repelled) on average from it. The presence of UDV generates random fluctuations of $\lambda_2(n)$ around zero, in such a way that we can study its statistical distribution \cite{kostelich}. 

We get a numerical approximation for the probability distribution of this exponent, denoted as $P(\lambda_2(n))$, by considering a large number of trajectories of length $n$ from initial conditions randomly chosen in the synchronization manifold. In Figure \ref{ftle}, we show some distributions of time-$50$ exponents, obtained for different values of the effective range $\alpha$, which here plays the role of a bifurcation parameter. The critical value for the onset of UDV is also the onset of intermittent transition to CS, at $\alpha = \alpha_c$, where an unstable-unstable pair bifurcation occurs. Numerically the onset of UDV was estimated through computing the value of $\alpha$ for which the fraction of positive finite-time Lyapunov exponents, $\phi(n) = \int_0^\infty P(\lambda_2(n)) d\lambda_2$, becomes non-zero, yielding the points on the curve $\epsilon_c(\alpha)$ depicted in Figure \ref{synchro}. Hence, for $\alpha < \alpha_c$, no shadowing breakdown {\it via} UDV is expected, and the chaotic synchronized trajectories obtained through the numerical solution of Eq. (1) are expected to be shadowed over a longer time interval \cite{shadows}, but which may be enough for practical purposes (e.g., when computing dimensions and entropies). 

\begin{figure}
\includegraphics[width=0.85\columnwidth,clip]{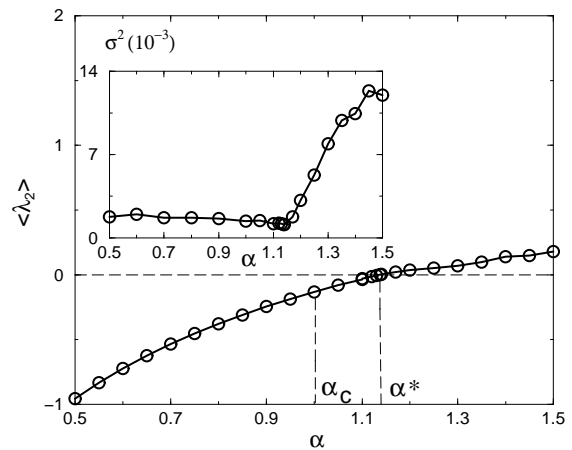}
\caption{\label{lyapmed}Average value and variance (inset) for the distribution of the finite-time Lyapunov exponents closest to zero {\it versus} $\alpha$, when $\epsilon = 1.0$}
\end{figure}

The shape of the probability distributions in Fig. \ref{ftle} is Gaussian-like, with different variances $\sigma^2$, according to the value which $\alpha$ takes on. The Gaussian nature of $P(\lambda_2(n))$ is already expected on general grounds \cite{kostelich}, and the distribution as a whole drifts toward positive values of $\lambda_2(n)$, as $\alpha$ increases. When $\alpha = \alpha^*$ the average $<\lambda_2(n)>$ crosses zero and have a maximum UDV (Fig. \ref{lyapmed}) since, at this point, $<\lambda_2(n)>$ equals the infinite-time limit of $\lambda_2(n)$, and UDV is more intense as the average of the time-n distribution crosses zero. This means that there are counterbalancing contributions from the negative and positive finite-time exponents, as illustrated by the distribution for $\alpha = 1.14834 \gae \alpha^*$ shown in Fig. \ref{ftle}. The point at which $<\lambda_2(n)> = 0$ also marks the loss of transversal stability of the synchronization manifold, or a {\it blowout bifurcation} \cite{ott}. Accordingly, the points on the curve $\epsilon^*(\alpha)$ (Fig. \ref{synchro}) were also computed by imposing that $<\lambda_2(n)> = 0$, which furnished the same results as those obtained with help of the order parameter ${\bar R}$, thus confirming the relationship between the loss of synchronization and the shadowing breakdown {\it via} UDV. 

In the vicinity of the point $\alpha = \alpha^*$ the shadowing times can be very short, and the validity of the computed numerical trajectories is doubtful. There results the time-$n$ exponents (with $n$ greater than the shadowing time), may suffer from similar shadowability problems, when taking individually, as the chaotic trajectories themselves. However, in terms of the numerical diagnostics of UDV, we are actually interested in statistical properties of the time-$n$ exponents, as their averages and variances. The former yields the point where UDV is the most intense, through a blowout instability, whereas the latter can be used to estimate shadowing times \cite{sauer}. On the other hand, in some physically relevant cases, statistical quantities like these have been found to be meaningful despite unstable dimension variability \cite{kurt}. 

For $\alpha > \alpha^*$, the relative number of  positive time-n exponents increases and we progressively return to a situation where UDV is less pronounced, and better shadowing properties are expected. For example, when $\alpha = 2.5$, Fig. \ref{ftle} indicates that almost all time-n exponents are positive. This means that, even though the lattice trajectories are far from being CS, they are nonetheless better shadowed than before. Figure \ref{lyapmed} also depicts the variance of the distributions, showing that it is roughly constant until $\alpha$ reaches the blowout value $\alpha^*$, after that the distribution broadens up and the variance increases.

In conclusion, we present numerical evidence that the intermittent transition to complete synchronization in a coupled map lattice is followed by a strong violation of hyperbolicity in the synchronization manifold. This breakdown begins with an unstable-unstable pair bifurcation, through which an unstable periodic orbit and all its pre-images embedded in the synchronization manifold lose transversal stability. This is the typical way in which invariant manifolds lose transversal stability in riddling-blowout bifurcation sequences. As a consequence of having been left with an infinite number of transversely stable orbits, the dimension of the unstable subspace varies from point to point along the synchronization manifold. There are severe consequences from the point of view of the validity of numerical trajectories obtained by computer simulation of high-dimensional extended systems, from which a coupled logistic map lattice has been chosen as a fundamental and representative example. The numerical trajectories may lose their validity, since they are no longer shadowed by true chaotic orbits for a reasonable time. The shadowing properties become worse as we approach the point whereby the synchronization manifold loses transversal stability. The statistical properties of transverse finite-time Lyapunov exponents indicate that both the completely synchronized and completely unsynchronized maps have adequate shadowing properties, and the validity of numerical results within those realms is fairly well established. The shadowing problems occur just at the transitional regime, where laminar regions of chaotic synchronized behavior coexist with irregular bursts of non-synchronized activity. While our discussion was based on a specific coupled map lattice, the general features obtained here are rather typical for complex systems presenting regular and chaotic behavior in space and time. The connection between shadowing breakdown and intermittent transition to complete synchronization identified in this work is a common feature of complex systems. 

This work was made possible through partial financial support from the following Brazilian research agencies: FAPESP, CNPq, CAPES, and Funda\c c\~ao Arauc\'aria. C.G. was also supported by A. V. Humboldt Foundation.

\end{document}